\documentclass[12pt]{JHEP3}

\usepackage[OT2,OT1]{fontenc}
\newcommand\cyr{%
\renewcommand\rmdefault{wncyr}%
\renewcommand\sfdefault{wncyss}%
\renewcommand\encodingdefault{OT2}%
\normalfont
\selectfont}
\DeclareTextFontCommand{\textcyr}{\cyr}
\usepackage{amsmath}
\usepackage{amsfonts}
\usepackage{amssymb}
\usepackage{graphicx}
\usepackage{bm}

\def\be{\begin{equation}}
\def\ee{\end{equation}}
\def\ba{\begin{eqnarray}}
\def\ea{\end{eqnarray}}
\def\bs{\begin{subequations}}
\def\es{\end{subequations}}
\newcommand{\bma}{\left(\begin{array}}
\newcommand{\ema}{\end{array}\right)}

\def\rme{e}
\def\rmd{d}

\def\p{\partial}

\def\cO{{\cal O}}
\def\cP{{\cal P}}
\def\cL{{\cal L}}
\def\cM{{\cal M}}
\def\N{\nabla}
\def\B{\Box}
\def\a{\alpha}
\def\k{\kappa}
\def\e{\epsilon}
\def\b{\beta}
\def\s{\sigma}

\newcommand{\Eq}[1]{(\ref{#1})}
\def\lp{\ell_{\rm Pl}}

\title{Cosmology of the Lifshitz universe}
\author{Gianluca Calcagni\\
Institute for Gravitation and the Cosmos, Department of Physics,\\ The Pennsylvania State University,\\
104 Davey Lab, University Park, Pennsylvania 16802, U.S.A.\\
E-mail: \email{gianluca@gravity.psu.edu}}

\date{April 6, 2009}

\abstract{We study the ultraviolet complete non-relativistic theory recently proposed by Ho\v{r}ava. After introducing a Lifshitz scalar for a general background, we analyze the cosmology of the model in Lorentzian and Euclidean signature. Vacuum solutions are found and it is argued the existence of non-singular bouncing profiles. We find a general qualitative agreement with both the picture of Causal Dynamical Triangulations and Quantum Einstein Gravity. However, inflation driven by a Lifshitz scalar field on a classical background might not produce a scale-invariant spectrum when the principle of detailed balance is assumed.}

\preprint{JHEP09(2009)112 \hspace{2cm} arXiv:0904.0829}
\begin{document}


\section{Motivation}

String theory \cite{Pol98,Zwi09}, loop quantum gravity \cite{thi01,AL,rov97} and spin-foams \cite{Per03} are among the most popular candidates for a theory of quantum gravity. Other independent approaches are asymptotically safe Quantum Einstein Gravity (QEG) \cite{Reu1}--\cite{BMS}
(reviewed in \cite{CPR,NiR,ReS4}) and Causal Dynamical Triangulations (CDT), a Lorentzian path-integral formulation of quantum gravity where the integral is performed over piecewise flat 4-geometries \cite{AmJ}--\cite{AGJL2}
(for a review consult \cite{lol08}). In conformity with the spirit of general relativy and quantum field theory, all these frameworks make the assumption that local Lorentz symmetry is exact at all scales. At sufficiently large scales, Lorentz invariance has been verified experimentally to a high degree of accuracy. However, at high energies Lorentz violation may occur without contradicting any observational constraint \cite{Mat05,JLM}. Some gravitational models are Lorentz invariant and implement a symmetry-breaking mechanism. Another legitimate possibility is that Lorentz invariance is not a fundamental property of Nature but an accidental symmetry of a low-energy theory \cite{Pav67,ChN}.

The latter perspective was instrumental for a recent proposal by Ho\v{r}ava, who constructed an ultraviolet (UV) complete theory of membranes \cite{Hor1} and gravity \cite{Hor2}. The problem was to find a $(D+1)$-dmensional quantum theory whose ground-state wavefunction reproduces the partition function of a given $D$-dimensional Euclidean (or, for curved backgrounds, Riemannian) theory. A physical example, mutuated from the theory of critical systems, which obeys this property is the Lifshitz scalar field \cite{Lif41a}--\cite{CL},
\be\label{f1}
S_{\rm Lifshitz}=\frac12\int \rmd t\rmd^D x\left[\dot\phi^2-\frac14(\Delta\phi)^2\right]\,,
\ee
where a dot denotes a derivative with respect to time $t$ and $\Delta=\p_i\p^i$ is the spatial Laplacian. The associated $D$-dimensional action can be shown to be
\be
W_{\rm Lifshitz}=\frac12\int \rmd^D x\,\p_i\phi\p^i\phi\,.\nonumber
\ee
Equation \Eq{f1} defines an anisotropic scaling between time and space, characterized by the dynamical critical exponent (or anisotropic scaling exponent) $z$ \cite{Car96}. In general anisotropic systems, coordinates scale as 
\be\label{as}
t\to b^z t\,,\qquad {\bf x}\to b{\bf x}\,,
\ee
for constant $b$, so that time and space have dimensions (in momentum units) $[t]=-z$ and $[x^i]=-1$; in this case, $z=2$. The two-point correlation function of the scalar field depends on the conformal dimension $[\phi]=(D-z)/2$, so that the critical exponent will determine the dimension $D$ at which the field propagator becomes logarithmic; when $D=z$, the system is said to be at (quantum) criticality. Reversing the logic, given a dimensionality $D$ the value of $z$ will characterize the critical behaviour of correlation functions near a phase transition. The meeting point of phase boundaries in multicritical phenomena is called multicritical point. For systems, such as certain metamagnets, liquid crystals and Ising models, displaying three phases (one disordered, one homogeneous, and one spatially uniform) the tricritical point is called \emph{Lifshitz point} \cite{Car96,CL,HLS}.

These models can be studied with renormalization group (RG) techniques \cite{Car96,fis99,BB,BTW,Pol01}. The Lifshitz scalar theory eq.~\Eq{f1} defines two Gaussian fixed points, one at $z=2$ where the system is invariant under the anisotropic scaling \Eq{as} and one at $z=1$ where the operator $(\Delta\phi)^2$ becomes irrelevant and local Lorentz invariance can be restored. Renormalizable theories with higher spatial derivatives were studied in \cite{AH,Ans1,Vis09} for scalars and fermions and in \cite{Ans2,Ans3} for gauge theories (see also \cite{Ans4} for a Lorentz-violating extension of the Standard Model).

The above construction was carried out in \cite{Hor1} for gravity at $z=2$ in $D+1$ dimension, with particular attention to the $(2+1)$-dimensional membrane theory at quantum criticality.\footnote{The main interest in this model lays in the fact that the ground state of a single membrane with a given compact topology reproduces, on one hand, the bosonic string partition function for the same worldsheet topology and, on the other hand, may pave the way for the construction of a many-membrane Fock space.} Later the same author proposed other gravitational theories with different dimensionality and value of the critical exponent: $z=3$ or $4$ in $3+1$ dimensions, $z=4$ in $4+1$ dimensions, and the ultralocal case $z=0$ \cite{Hor2}.

The notion of dimension at short scales is one of the quantum properties of `geometry' which may radically differ from the classical macroscopic picture. In these cases, spacetime is said to `emerge' from ultraviolet physics. In some approaches to quantum gravity the dimensionality of spacetime is defined by standard tools of fractal geometry \cite{Fal03} and the ordinary topological dimension is recovered at large scales/low energy.

For example, the Hausdorff dimension has been employed in $2D$ Euclidean quantum gravity \cite{ADJ}--\cite{ABNRW},
$4D$ gravity \cite{AMM} and CDT \cite{AJL5}. On the other hand, the \emph{spectral dimension}\footnote{Introductions on the subject can be found in \cite{AJW,ABNRW,bH}.} is particularly suitable for capturing the fractal behaviour of $2D$ Euclidean quantum gravity \cite{AJW,ABNRW,KN,JW98,CW} (see also \cite{DD}--\cite{JS} 
 for some formal studies on random geometrical objects), CDT \cite{AJL4,AJL5}, asymptotically safe gravity \cite{LaR5}, LQG and spin-foams \cite{Mod08}.

Ho\v{r}ava's theory with $z=3$ shares a remarkable property with CDT, QEG and spin-foams. Namely, near the Planck scale gravity feels only two of the four spacetime dimensions or, more precisely, the spectral dimension of the universe at small scales is 2 \cite{Hor3}.\footnote{Based on scaling properties of the area operator, it was shown that the spectral dimension of spacetime in LQG is 2.5 at Planck scale and 3 in the deep UV \cite{Mod08} (it is 3 also for $\kappa$-Minkowski \cite{Ben08}). The spectral dimension of the spatial section of LQG has been calculated in the kinematical Hilbert space of the theory, and it is not strictly related to the dimension of physical spacetime. Therefore the discrepancy at small scales between this result and those of CDT, QEG, spin-foams and the anisotropic theory may be a kinematical effect \cite{mod09}.} This is true for any $z=D$ critical theory, since in that case the propagator of the graviton or a scalar is logarithmic. This feature is of course no coincidence and it shows how a short-scale two-dimensional behaviour of Nature be essential in most UV finite models of gravity (in string theory this is true by construction). 

Here we wish to draw further comparison, focussing on vacuum cosmology, between the $z=3$ critical theory on one hand and CDT and QEG on the other. In CDT, at large-enough scales the Euclidean universe can be described by a de Sitter geometry perturbed by semi-classical fluctuations \cite{AGJL1,AJL7,AGJL2}. The characteristic size of the universe is roughly between $\lp$ and $O(10)\lp$, indicating that a semi-classical minisuperspace approximation, based on an FRW metric with positive spatial curvature, may be a fair description of the very early universe. However, near Planck energy geometry deviates from a smooth one, thus displaying fractal behaviour. The cosmology of QEG is again asymptotically de Sitter and it has a big bang singularity, perhaps as a consequence of the Einstein--Hilbert truncation \cite{BoR1,ReS3,BoR3}. In both cases, a semi-classical description of the universe breaks down near the big bang and a transition to a full quantum regime takes place. On the other hand, Ho\v{r}ava's proposal of trading exact Lorentz invariance for anisotropic scaling seems capable of describing some of the fractal properties of the deep quantum region even within a classical formalism. Clearly all these models should agree qualitatively, as they are all based on the classical Einstein--Hilbert action or its modifications. When matter is taken into account, however, viable inflation in a classical Lifshitz universe is difficult to achieve and a full RG analysis might be required.

To summarize, the aims and results of the present work are:
\begin{itemize}
\item To introduce scalar matter in Ho\v{r}ava's $z=3$ theory, thus extending the original proposal for pure gravity. \textit{We assume detailed balance}, which is not an indispensable ingredient of the theory but it makes its quantum properties simpler to analyze. Therefore we start from a three-dimensional action with non-local pseudo-differential operators. Under a `separate' detailed balance condition, one obtains a minimally coupled four-dimensional $z=3$ Lifshitz scalar action.
\item Study the cosmology of the model, with and without matter. We find vacuum solutions and argue that bouncing solutions exist and avoid the big bang singularity.
\item Solutions with Euclidean signature are asymptotically de Sitter and in qualitative agreement with the CDT scenario. The correspondence with CDT, already noticed in \cite{Hor2}, is supported now at the level of mini-superspace.
\item On the other hand, inhomogeneous scalar perturbations against a classical background, generated by quantum fluctuations of an inflationary Lifshitz field, are unable to yield a scale-invariant spectrum. Abandoning the detailed balance condition one can obtain scale invariance. This result is not obvious in the vacuum theory and suggests to modify the original formulation.
\end{itemize}
The paper is organized as follows. The gravitational sector is reviewed in section \ref{gra} and Lifshitz matter is introduced in section \ref{lifss}. Section \ref{cosmo} is devoted to the cosmological properties of the model: vacuum solutions of universes with Lorentzian and Euclidean signature are described in sections \ref{lor} and \ref{rie}. Cosmological perturbations and the inflationary spectrum are discussed in section \ref{pert}.


\section{Action}\label{acti}

Let $\cM=\mathbb{R}\times\Sigma$ be a time-space manifold with signature $({-},{+},{+},{+})$ embedding a torsion-free three-dimensional space $\Sigma$ with dimensionless metric $g_{ij}$, where Latin indices run from 1 to 3. On $\Sigma$ we define the space-covariant derivative on a covector $v_i$ as $\N_i v_j \equiv \p_i v_j-\Gamma^l_{ij}v_l$, where $\Gamma^l_{ij}\equiv g^{lm}\left[\p_{(i} g_{j)m}-\tfrac12\p_m g_{ij}\right]$ is the spatial Christoffel symbol. The curvature invariants (under spatial diffeomorphisms) quadratic in spatial derivatives of the metric are the Riemann tensor $R^l_{~imj}\equiv \p_m \Gamma^l_{ij}-\p_j \Gamma^l_{im}+\Gamma^n_{ij}\Gamma^l_{mn}-\Gamma^n_{im}\Gamma^l_{jn}$, the Ricci tensor $R_{ij}\equiv R^l_{~ilj}$ and the Ricci scalar $R\equiv R_{ij}g^{ij}$.


\subsection{Gravity}\label{gra}

Given these definitions, the Ho\v{r}ava $3+1$ action with $z=3$ is \cite{Hor2}
\be\label{act}
S_g=\int_\cM\rmd t\rmd^3x\, \sqrt{g}\,N(\cL_K-\cL_V)\,,
\ee
where $g$ is the determinant of the 3-metric and $N=N(t)$ is a dimensionless homogeneous gauge field. The kinetic term is
\be\label{kin}
\cL_K=\frac{2}{\k^2}\cO_K=\frac{2}{\k^2}\left(K_{ij}K^{ij}-\lambda K^2\right)\,,
\ee
where $\k^2$ and $\lambda$ are coupling constants with dimension $[\k^2]=z-3$ and $[\lambda]=0$ (hence both dimensionless at the $z=3$ Lifshitz point), $K_{ij}=K_{ij}(t,{\bf x})$ is
\be\label{K}
K_{ij}=\frac1N \left[\frac12\dot g_{ij}-\N_{(i}N_{j)}\right]\,,
\ee
and $K\equiv K_i^{\ i}$. Here $N_i=N_i(t,{\bf x})$ is a gauge field with scaling dimension $[N_i]=z-1$ and round brackets denote symmetrized indices, $X_{(ij)}=\left(X_{ij}+X_{ji}\right)/2$. Eq.~\Eq{kin}, once generalized to arbitrary dimension $D$, is the most general kinetic term invariant under foliated diffeomorphisms \cite{Hor1,Hor2}.

The `potential' term $\cL_V$ of the $(D+1)$-dimensional theory is determined by the \emph{principle of detailed balance} \cite{Car96}, requiring $\cL_V$ to follow, in a precise way, from the gradient flow generated by a $D$-dimensional action $W_g$. This principle was applied to gravity \cite{Hor1,Hor2}, with the result that the number of possible terms in $\cL_V$ are drastically reduced with respect to the broad choice available in an effective field theory. Below we shall illustrate how it works in the scalar sector. For pure gravity, the most general covariant Riemannian action in three dimensions with $z=3$ anisotropy and all possible relevant operators is \cite{Hor2}
\be
W_g= \frac{1}{\nu^2}\int \omega_3(\Gamma)+\mu\int\rmd^3 x \sqrt{g}\left(R-2\Lambda_W\right)\,,
\ee
where $\omega_3$ is the Chern--Simons form and $\nu$, $\mu$ and $\Lambda_W$ are real constants with dimension $[\nu]=0$, $[\mu]=1$ and $[\Lambda_W]=2$, respectively. The associated spacetime `potential' is
\ba
\cL_V&=&\sum_{A=2}^6 (-1)^{A}\a_A\cO_A\nonumber\\
     &=& \a_6 C_{ij}C^{ij}-\a_5 \e^{ij}_{\ \ l}R_{im}\N_j R^{ml}+\a_4\left[R_{ij}R^{ij}-\frac{4\lambda-1}{4(3\lambda-1)}R^2\right]\nonumber\\
     &&+\a_2(R-3\Lambda_W)\,,
\ea
where $A=[\cO_A]$ is the number of spatial derivatives of the metric and
\be
C^{ij}\equiv \e^{ilm}\N_l\left(R_m^j-\frac14\delta_m^jR\right)
\ee
is the Cotton tensor \cite{Hor2} and $\e^{ilm}$ is the Levi--Civita symbol. Up to a total derivative (which we discard for simplicity together with any other boundary term) and making use of the twice-contracted Bianchi identity, the 6th-order operator can be written as
\be
\cO_6= \frac18R\Delta R -R_{jl} \Delta R^{jl}+R_{jl} \N_i\N^j R^{il}\,,
\ee
where $\Delta\equiv \N_i\N^i$. The coupling constants are
\bs\ba
\a_6 &=& \frac{\k^2}{2\nu^4}\,,\qquad \a_5 = \frac{\k^2\mu}{2\nu^2}\,,\\
\a_4 &=& \frac{\k^2\mu^2}{8}\,,\qquad \a_2 = \frac{\a_4\Lambda_W}{3\lambda-1}\,,
\ea\es
and have dimension $[\a_A] = z+3-A$.

The action \Eq{act} contains higher-order spatial derivatives of the metric but is second-order in time derivatives; hence there are no ghosts, if $\lambda$ is chosen to yield the correct sign in front of the kinetic term (this is the case for $\lambda>1/3$). Also, it is invariant under foliated diffeomorphisms, i.e., diffeomorphisms preserving the codimension-one foliation of $\cM$ \cite{MM} with leaves $\Sigma$. Foliated diffeomorphisms consist in time-dependent time reparametrizations and spacetime-dependent spatial diffeomorphisms; hence, they are generated by the infinitesimal transformations
\be
t\to t+f(t)\,,\qquad x^i\to x^i+\zeta^i(t,{\bf x})\,.
\ee
In second-order Hamiltonian framework, to each of the six gauge symmetries there corresponds a first-class constraint involving the nine canonical variables $N^i$ and $K_{ij}$ and their conjugate momenta; the total number of degrees of freedom is therefore three, one more than in general relativity and given by the trace of graviton modes \cite{Hor1}. When $\lambda=1/3$ the system acquires another gauge symmetry, i.e., invariance under local conformal transformations of the metric \cite{Hor2}, by virtue of which the extra degree of freedom is gauged away. This happens also when $\lambda=1$, where the infrared (IR) theory is invariant under full spacetime diffeomorphisms. In that case there is an extra diffeomorphism symmetry and the gauge fields $N$, $N^i$ and $g_{ij}$ are interpreted, respectively, as the ADM lapse, shift and spatial components of the four-dimensional Lorentzian metric $g_{\mu\nu}$. In particular, eq.~\Eq{K} is the ADM extrinsic curvature \cite{wal84}.

Near the Lifshitz point $z=3$, the operator $\cO_6$ and the kinetic term are marginal, while the other operators are relevant. These induce a flow from the ultraviolet fixed point at $z=3$ to the infrared fixed point at $z=1$, where the only relevant operators are the kinetic term and $\cO_2$ (then $\cO_4$ becomes marginal and the other operators irrelevant):
\be\label{irs}
S_g\sim\frac{2}{\k^2}\int\rmd t\rmd^3x\,\sqrt{g}\,N\left[K_{ij}K^{ij}-\lambda K^2+c^2(R-3\Lambda_W)+\left(\frac{\k^2\mu}{4}\right)^2\cO_4\right]\,,
\ee
where
\be
c\equiv\sqrt{-\frac{\k^2\a_2}{2}}=\frac{\k^2\mu}{4}\sqrt{\frac{\Lambda_W}{1-3\lambda}}.
\ee
After rescaling $t\to t/c$ (i.e., defining $x^0=ct$ as the new time variable), $N_i\to cN_i$ and defining the effective cosmological constant $\Lambda=3\Lambda_W/2$ and the effective Newton's constant
\be
G\equiv\frac{\k^2}{32\pi c}\,,
\ee
eq.~\Eq{irs} reads, up to the $\cO_4$ term,
\be
S_g\sim\frac{1}{16\pi G}\int\rmd^4 x\,\sqrt{g}\,N\left(K_{ij}K^{ij}-\lambda K^2+R-2\Lambda\right)\,,
\ee
which coincides with the Einstein--Hilbert action with cosmological constant in the limit $\lambda\to 1$. Remarkably \cite{Hor2}, the gravitational constant $G$, speed of light $c$ and cosmological constant all stem from the couplings of operators relevant at the UV fixed point, and they have the correct scaling dimension in the infrared ($[G]=-2$, $[c]=z-1$). Note, however, that in the relativistic limit the cosmological constant is negative definite in order for the resulting theory to be Lorentzian ($c$ real). In particular, de Sitter and Minkowski are not vacuum solutions.


\subsection{Scalar matter}\label{lifss}

At this point we would like to add a matter sector with the following properties: It must (i) respect foliated diffeomorphism invariance, (ii) obey the principle of detailed balance and (iii) be nontrivial at the $z=3$ critical point and Lorentz invariant in the infrared. All these are defining properties of the theory, although any or all of them might be relaxed in effective or general UV-finite models (e.g., see \cite{Vis09}). Below we shall slightly relax (ii), although the requirements (i) and (iii) will be strictly enforced.

Here we consider a `Lifshitz' scalar for $z=3$ anisotropic scaling. The aim is to find a three-dimensional covariant Riemannian action $W_\phi$ such that it exhibits $z=3$ anisotropic scaling and the spacetime four-dimensional action of the scalar field be
\be\label{sca1}
S_{\phi}=\frac{1}{2}\int \rmd t\rmd^3x \sqrt{g}N\left[\frac{3\lambda-1}{2}\frac{\dot{\Phi}^2}{N^2}-\left(\frac{\delta W_\phi}{\delta\phi}\right)^2\right]\,,
\ee
where $\dot{\Phi}\equiv\dot\phi-N^i\p_i\phi$. The $\lambda$-dependent factor in front of the kinetic term is for later convenience.

Here we are making an assumption we should immediately stress. Proper implementation of the detailed balance principle would require to define a `metric of fields' $\mathbb{G}$ incorporating both the generalized DeWitt metric of metrics ${\cal G}$ \cite{Hor1,Hor2} and the scalar-field component. Let us choose a diagonal metric and matrix field
\[
\mathbb{G}=\bma{cc} {\cal G}/g &\,\, 0 \\ 0 &\,\, 1/g\ema\,,\qquad
q=\bma{cc} g_{ij}\,\, & 0 \\ 0 &\,\, \phi\ema\,.
\]
The potential term of the total $(3+1)$-dimensional action $S=S_g+S_\phi$ should be defined, symbolically, as 
\be\label{sb1}
{\rm tr} \left(\frac{\delta W}{\delta q}\mathbb{G}\frac{\delta W}{\delta q}\right)\,,
\ee
where $W=W_g+W_\phi$. However, even choosing a $W_\phi$ with minimal coupling the scalar field in $S$ would be non-minimally coupled through $\delta W/\delta g_{ij}$ contributions. To avoid this complication, we content ourselves with the much milder `separate' detailed balance encoded in eq.~\Eq{sca1}, corresponding to the replacement
\be\label{sb2}
\frac{\delta W}{\delta q}=\bma{cc} \frac{\delta W}{\delta g_{ij}} &\,\, 0 \\ 0 &\,\, \frac{\delta W}{\delta\phi}\ema\to
\bma{cc} \frac{\delta W_g}{\delta g_{ij}} &\,\, 0 \\ 0 &\,\, \frac{\delta W_\phi}{\delta\phi}\ema.
\ee
To a certain degree, this will affect the inheritance of quantum properties of the theory $S$ from the lower-dimensional theory $W$ but the above operation guarantees a simpler spacetime action which will suffice for our purposes.

The Riemannian action $W_\phi$ does not feature ordinary operators because of the requirement of $z=3$ anisotropy. The only UV marginal operator in $W_\phi$ whose variation and square gives the 6th-order operator $\phi\Delta^3\phi$ is $\phi\Delta^{3/2}\phi$, where $\Delta^{3/2}$ is a pseudo-differential operator \cite{Gil95,Gru96}. Our ansatz for $W_\phi$
is
\be\label{wphi}
W_{\phi}=\frac{1}{2}\int \rmd^3x \sqrt{g}\left[-\s_3\phi\Delta^{3/2}\phi-\s_2\phi\Delta\phi
+m\phi^2\right]\,,
\ee
where the coupling constants (all assumed to be positive) have scaling dimension $[\s_i]=z-i$ and $[m]=z$. One could also allow for a more general potential $U(\phi)$ but it would proliferate the number of operators in the spacetime action. Therefore we shall keep only the mass term, which is necessary to restore Lorentz invariance in the IR limit.

Pseudo-differential operators of the type $\Delta^\a$ and $\B^\a$ for arbitrary $\a$, of which fractional derivatives are a subset, have been receiving much attention and there exists a fairly wide dedicated literature \cite{See67}--\cite{BK2}
(and references therein). Lorentz-invariant non-local theories with fractional differential operators lead to qualitatively different conceptual frameworks with respect to standard classical and quantum field theory. For instance, Huygens' principle is violated (obeyed) in even (respectively, odd) spacetime dimensions \cite{Gia91,BGO,BG}. Nevertheless, definitions of these operators, formal solutions of nonlocal equations, quantization and causality are all well-established \cite{BG,BBOR}, also in Euclidean theories \cite{BGO,BG,Lim06}.

Taking the functional derivative of eq.~\Eq{wphi} with respect to $\phi$, we get\footnote{Integration by parts of fractional pseudo-differential operators $F(\Delta)$ or $F(\B)$ may be intuitively understood as follows. One assumes that $F$ admits a series representation of the form $F(\B)=\sum_n a_n \B^n$, where $n\in\mathbb{N}$ and $a_n$ are the Taylor coefficients of $F$. From this definition, most of the properties of $F$ (including chain rule and integration by parts) naturally reproduce those of ordinary differential operators. Unfortunately, in most of the cases the series representation is only a formal tool, because either it does not converge on the chosen Hilbert space or it is not even well-defined to begin with. In the former case one can count the exponential operator $\rme^\B$, which plays an important role in string field theory. There, the coefficients $a_n=1/n!$ are well-defined, but when one applies the operator to a test function the series will not converge generally; so one must resort to a different representation (for example the one in terms of the heat kernel \cite{cuta2}--\cite{cuta5}).
On the other hand, the latter case is epitomized by the square root of the Beltrami--Laplace operator. One can define the operator $\sqrt{\B+\varepsilon}$ as a binomial series, perform any operation formally, and finally take the limit $\varepsilon\to 0$ after resumming \cite{doA92}. This problem is bypassed by taking a suitable integral representation \cite{BG,BBOR}, which coincides with the naive one at formal level.}
\be
\frac{1}{\sqrt{g}}\frac{\delta W_\phi}{\delta\phi}= -\s_3\Delta^{3/2}\phi-\s_2\Delta\phi+m\phi\,.
\ee
Then,
\be
S_{\phi}=\frac{1}{2}\int \rmd t\rmd^3x \sqrt{g}N\left[\frac{3\lambda-1}{2}\frac{\dot{\Phi}^2}{N^2}-\sum_{A=2}^6\b_A\cP_A-m^2\phi^2\right]\,,
\ee
where
\be
\cP_A=\phi\Delta^{A/2}\phi\,,
\ee
$[\cP_A]=3+A-z$ and
\ba
\b_6&=& \s_3^2\,,\qquad \b_5=2\s_3\s_2\,,\qquad \b_4= \s_2^2\,,\\
\b_3&=&-2\s_3m\,,\qquad \b_2= -2\s_2m\,.
\ea
At the UV fixed point matter behaves as a $z=3$ Lifshitz scalar,
\be
S_{\phi}\sim\frac{1}{2}\int \rmd t\rmd^3x \sqrt{g}N\left[\frac{3\lambda-1}{2}\frac{\dot{\Phi}^2}{N^2}-\b_6\phi\Delta^3\phi\right]\,.
\ee
Relevant deformations then push the system towards the IR fixed point, where Lorentz invariance is restored:
\be
S_{\phi}\sim\frac{1}{2}\int \rmd t\rmd^3x \sqrt{g}N\left[\frac{3\lambda-1}{2}\frac{\dot{\Phi}^2}{N^2}-|\b_2|\p_i\phi\p^i\phi-m^2\phi^2\right]\,.
\ee
The operators in the total action $S=S_g+S_\phi$ are summarized in table \ref{tab1}.
\begin{table}
\begin{tabular}{|l|c||c|c|c|}\hline
$\cO$       			 & $[\cO]$ &   $z=3$ (UV fixed point) &    $z=2$    &  $z=1$ (IR fixed point) \\ \hline\hline
$\cO_K$   			   &   $2z$  &   marginal               &  relevant   & relevant   \\
$\cO_6$   			   &     6   &   marginal               &  irrelevant & irrelevant \\
$\cO_5$   			   &     5   &   relevant               &  marginal   & irrelevant \\
$\cO_4$   			   &     4   &   relevant               &  relevant   & marginal   \\
$\cO_2$   			   &     2   &   relevant               &  relevant   & relevant   \\\hline
$\dot\Phi^2$			 &   $3+z$ &   marginal               &  marginal   & marginal   \\
$\cP_6$            &   $9-z$ &   marginal               &  irrelevant & irrelevant \\
$\cP_5$            &   $8-z$ &   relevant               &  irrelevant & irrelevant \\
$\cP_4$            &   $7-z$ &   relevant               &  marginal   & irrelevant \\
$\cP_3$            &   $6-z$ &   relevant               &  relevant   & irrelevant \\
$\cP_2$         	 &   $5-z$ &   relevant               &  relevant   & marginal   \\
$\phi^2$           &   $3-z$ &   relevant               &  relevant   & relevant   \\\hline
\end{tabular}
\caption{\label{tab1} Summary of the operators $\cO$ in the four-dimensional action and their properties under renormalization group flow from $z=3$ (UV) to $z=1$ (IR).}
\end{table}


\subsection{Equations of motion}

Variation of the total action with respect to $N$ yields
\be\label{neom}
\frac{2}{\k^2}\left(\lambda K^2-K_{ij}K^{ij}\right)-\cL_V=\rho\,,
\ee
where
\be
\rho\equiv-\frac{1}{\sqrt{g}}\frac{\delta S_\phi}{\delta N}=\frac{1}{2}\left[\frac{3\lambda-1}{2}\frac{\dot{\Phi}^2}{N^2}+\left(\frac{\delta W_\phi}{\delta\phi}\right)^2\right]\,,\qquad [\rho]=z+3\,.
\ee
The equation of motion of the scalar field $\delta S/\delta\phi=0$ is, in the $N^i=0$ gauge,
\be\label{feom}
\frac{3\lambda-1}{2}\frac{1}{N\sqrt{g}}\p_t\left(\frac{\sqrt{g}\dot\phi}{N}\right)+\sum_A\b_A\Delta^{A/2}\phi+m^2\phi=0\,.
\ee


\section{Cosmology}\label{cosmo}

We now specialize to a Friedmann--Robertson--Walker (FRW) background. In synchronous time $t$, the cosmological ADM metric has $N=1$, $N_i=0$ and $g_{ij}=a(t)^2\tilde g_{ij}$, where $a(t)$ is the scale factor and
\be
\tilde g_{ij}\rmd x^i \rmd x^j=  \frac{\rmd r^2}{1-\textsc{k}\,r^2}+r^2(\rmd\theta^2+\sin^2\theta\rmd\varphi^2)
\ee
is the line element of the maximally symmetric three-dimensional space $\tilde\Sigma$ of constant sectional curvature $\textsc{k}$ (equal to $-1$ for an open universe, 0 for a flat universe and $+1$ for a closed universe with radius $a$). On this background,
\be
K_{ij}=\frac{H}{N}g_{ij}\,,\qquad R_{ij}=\frac{2\textsc{k}}{a^2}g_{ij}\,,\qquad C_{ij}=0\,,
\ee
where $H\equiv \dot a/a$ is the Hubble parameter and we have exploited the symmetries of $\tilde\Sigma$ \cite{wei72}.
The minisuperspace action reads
\ba
S_{\rm FRW}&=&\int\rmd t\rmd^3x\, a^3N\,\left[\frac{3(1-3\lambda)}{16\pi Gc}\frac{H^2}{N^2}-\frac{\a_4}{1-3\lambda}\left(\frac{\textsc{k}}{a^2}\right)^2-6\a_2\left(\frac{\textsc{k}}{a^2}-\frac{\Lambda}{3}\right)\right]\nonumber\\
&&+\frac12\int\rmd t\rmd^3x\, a^3N\,\left[\frac{3\lambda-1}{2}\frac{\dot\phi^2}{N^2}-m^2\phi^2\right]\,.\label{mini}
\ea


\subsection{Lorentzian cosmology}\label{lor}

The case $\lambda>1/3$ corresponds to a negative cosmological constant. The first Friedmann equation \Eq{neom} is
\be\label{feq}
H^2=\frac{8\pi \tilde Gc}{3}\rho-\frac{B^2}{a^4}-\frac{c^2\tilde{\textsc{k}}}{a^2}-\frac{c^2|\tilde\Lambda|}{3}\,,
\ee
where tilded quantities are 
\be
\tilde{\textsc{k}}=\frac{2\textsc{k}}{3\lambda-1}\,,\qquad \tilde\Lambda=\frac{3\Lambda_W}{3\lambda-1}\,,\qquad \tilde G= \frac{2G}{3\lambda-1}\,,
\ee
and
\be
B=\frac{\k^2\mu\tilde{\textsc{k}}}8\,.
\ee
The contribution $-a^{-4}$ is reminiscent of the dark radiation term in braneworld cosmology \cite{BDEL} and, notably, the effective early-time energy density of the cosmological condensate of \cite{AlB,ABC} in the strong coupling limit.

The Klein--Gordon equation \Eq{feom} is 
\be
\ddot\phi+3H\dot\phi+\frac{2m^2}{3\lambda-1}\phi=0.
\ee
The second Friedmann equation is obtained by varying the total minisuperspace action with respect to $a$ or by deriving eq.~\Eq{feq} and plugging the Klein--Gordon equation in:
\be\label{feq2}
\frac{\ddot a}{a}=-\frac{4\pi \tilde G c}{3}(\rho+3p)+\frac{B^2}{a^4}-\frac{c^2|\tilde\Lambda|}{3}\,,
\ee
where $p=\cL_\phi$ is the scalar field pressure.

In vacuum ($\rho=0=p$) the first Friedmann equation is well-defined only if the universe is open ($\tilde{\textsc{k}}<0$; $\tilde{\textsc{k}}=-1$ in what follows). The only vacuum solution is a static ($a=1$) anti-de Sitter universe with $|\tilde\Lambda|=3(1-B^2/c^2)$.

In the presence of matter, at the critical time $t_*$, $a=a_*$, $H_*=0$ and the universe undergoes a bounce. This happens 
when
\be
\rho=\rho_*= \frac{c|\Lambda|}{8\pi G}+\frac{3c\textsc{k}}{8\pi G a_*^2}\left(1+\frac{B^2}{a_*^2c^2\tilde{\textsc{k}}}\right)\,.
\ee
The critical energy density $\rho_*$ is determined by the couplings of the theory, and increases for an open, flat ($\tilde{\textsc{k}}=0$) and closed ($\tilde{\textsc{k}}>0$) universe. In a quasi-de Sitter regime (slow rolling, $|\dot\phi|\ll |m\phi|$), this becomes a lower bound on the field expectation value $\phi$. For instance, for a flat universe and a quadratic potential
\be
\phi_*= \pm\sqrt{\frac{c|\Lambda|}{4\pi G m^2}}\,.
\ee
From an inspection of the equations of motion \Eq{feq} and \Eq{feq2}, one can see that not all matter contents will admit a bounce or even a global solution (for instance, in the presence of radiation only). It would be interesting to find explicit solutions realizing bouncing scenarios, where the big bang singularity is avoided.


\subsection{Riemannian vacuum solutions and $\lambda<1/3$}\label{rie}

It is worth mentioning that the Wick-rotated theory admits cosmological vacuum solutions. The Friedmann equations are (tilde's omitted)
\ba
H^2&=&\frac{8\pi Gc}{3}\rho_{\rm E}+\frac{B^2}{a^4}+ \frac{c^2\textsc{k}}{a^2}+ \frac{c^2|\Lambda|}{3}\,,\label{feqe}\\
\frac{\ddot a}{a}&=&-\frac{4\pi G c}{3}(\rho_{\rm E}+3p_{\rm E})-\frac{B^2}{a^4}+\frac{c^2|\Lambda|}{3}\,,
\ea
where a subscript E denotes Euclidean matter. Ignoring the latter, the general solution is
\be
a(t)_\pm=\frac{1}{2\sqrt{|\Lambda|}}\sqrt{\frac{1}{c^2}\rme^{\pm 2\sqrt{\frac{|\Lambda|}{3}}ct}+3(3c^2\textsc{k}^2-4B^2|\Lambda|)\rme^{\mp 2\sqrt{\frac{|\Lambda|}{3}}ct}-6\textsc{k}}\,.
\ee
The flat case is the de Sitter solution. From eq.~\Eq{feqe}, $H=0$ when the scale factor achieves the value
\be
a_*^\pm=\sqrt{\frac{3\left[-\textsc{k}\pm\sqrt{\textsc{k}^2-4|\Lambda|B^2/(3c^2)}\right]}{2|\Lambda|}}\,.
\ee
For an open universe, there is the possibility of a bounce. Let us take the solution $a_+$ with $\textsc{k}=-1$. A bounce does occur if $3c^2\textsc{k}^2-4B^2|\Lambda|>0$, the Hubble parameter always increasing (superacceleration); asymptotically the universe is de Sitter. If $3c^2\textsc{k}^2-4B^2|\Lambda|=0$, the universe evolves monotonically from Euclidean vacuum $a\sim \sqrt{3/(2|\Lambda|)}$ to de Sitter. If $3c^2\textsc{k}^2-4B^2|\Lambda|<0$, the universe evolves monotonically from a singularity to de Sitter, passing from a phase of normal expansion to acceleration to superacceleration. Notice that detailed balance realizes the second case, but the others are possible if one relaxes the conditions imposed on the coupling constants.

A closed universe, too, is asymptotically de Sitter, but it has a big bang singularity. This is in agreement with the CDT and QEG approaches.

Finally, we observe that the $\lambda<1/3$ case in \emph{Lorentzian} signature (positive $\Lambda$) mimics Euclidean cosmology, as the kinetic terms change sign. Then, up to the $B$ term and a minus sign, eq.~\Eq{mini} in the absence of matter is the minisuperspace FRW action $S_{\rm E}$ in Euclidean signature. This is the same effective action found in CDT at large scales \cite{AJL5,AGJL1}: 
\[
S_{\rm CDT}=-S_{\rm E}\sim S_{\rm FRW}(\lambda<1/3)\,.
\]
When one inserts the scalar field in the $\lambda<1/3$ action,
\be
S_{\rm FRW}\left(\lambda<\frac13\right)\sim -\left[S_{\rm E}+\frac12\int \rmd t\rmd^3 x\,a^3\left(\frac{1-3\lambda}{2}\dot\phi^2+m^2\phi^2\right)\right]\,.
\ee


\subsection{Inflationary perturbations}\label{pert}

As long as $\lambda>1/3$, at large scales the cosmology of section \ref{lor} is the same as that of general relativity for any value of $\lambda$. The symmetry reduction to an FRW background drops the operators $\cO_6$ and $\cO_5$ which are, respectively, marginal and relevant at the UV fixed point. This occurs because the FRW background is insensitive of the anisotropic scaling. It is therefore natural to probe the physics of the `Lifshitz universe' at early times, when inhomogeneous perturbations are produced by quantum fluctuations of the inflaton field $\phi$.

To have a qualitative picture and obtain the greatest deviation from the standard scenario, we shall concentrate on the UV marginal operators. We implicitly set $\lambda=1$, in which case the gravitational sector exhibits only two degrees of freedom, i.e., the two polarization modes of the graviton represented by the transverse-traceless tensor $h_{ij}$ \cite{Hor1,Hor2}. The action, quadratic in tensor perturbations, was calculated in \cite{Hor1,Hor2} for a Minkowski background. For a flat FRW background in conformal time $\tau$ ($N=a$) it is easy to show that
\be
\delta^{(2)}S_g=-\frac{1}{2\k^2}\int \rmd \tau\rmd^3x\, a^2\left[h^{ij} h_{ij}''-\left(\frac{\k^2}{2\nu^2}\right)^2a^2h_{ij}\Delta^3h^{ij}\right]\,,
\ee
where primes are conformal derivatives. Denoting as $h_k$ the Fourier mode with wavenumber $k$ of one polarization and after a standard variable redefinition $v_k=ah_k$, the equation of motion of tensor perturbations in momentum space ($\Delta\to -k^2/a^2$) is
\be\label{h}
v_k''+\left[\left(\frac{\k^2}{2\nu^2}\right)^2\frac{k^6}{a^4}-\frac{a''}{a}\right]v_k=0\,.
\ee
Similarly, the perturbed Klein--Gordon equation for a test scalar field $u_k=a\delta\phi_k$ is
\be\label{df}
u_k''+\left[-\s_3^2\frac{k^6}{a^4}-\frac{a''}{a}+m^2\right]u_k=0\,.
\ee
Here we have neglected the backreaction of the metric, which would modify only the effective mass term. Notice that the 6th-order term in eq.~\Eq{df} is quadratic in the three-dimensional coupling $\sigma_3$, whose sign therefore is unimportant.

The fields in eqs.~\Eq{h} and \Eq{df} obey a particular case of the generalized Corley--Jacobson dispersion relation \cite{MB1}. This has been the subject of intensive study in the context of trans-Planckian cosmology and the ensuing spectra are well known \cite{MB1}--\cite{MB3}. Let 
\be
w_k''+\left[k^2\pm(Ck)^2\left(\frac{k^2}{a^2}\right)^{z-1}-\frac{a''}{a}\right]w_k=0
\ee
be the equation of motion of a gauge-invariant perturbation $w_k$ (a scalar or tensor mode) with effective squared mass $m_{\rm eff}^2\sim a''/a$. Here $C>0$, $z>1$ and we have included also the contribution of the relevant operator $w\Delta w$. For a power-law scale factor $a=|\tau|^p$, the above equation can be solved exactly. For the $+$ sign choice or when $C=0$, the cosmological spectrum (two-point correlation function) is scale invariant in a quasi-de Sitter regime ($p\lesssim-1$):
\be\label{specp}
k^3P_w=k^3 \frac{|w_k|^2}{a^2}= k^{2(1+p)}\,,\qquad n_w\equiv \frac{\rmd\ln (k^3P_w)}{\rmd\ln k}=2(1+p)\,,
\ee
where we introduced the spectral index $n_w\ll 1$. In particular, the tensor spectrum is scale invariant. For the $-$ sign choice, however, the spectrum is \cite{MB1,BM1}
\be\label{spec}
k^3P_w= k^{2(1+p)} \rme^{A k^z}\cos^2\left[2\pi|p|-\frac{\pi}{4}-C^{\frac{1}{p(z-1)}} k^{1+\frac{1}{p}}\right]\,,
\ee
where $A\gg 1$ for wavenumbers $k\sim 2\pi$. Let us ignore the oscillatory contribution. The exponential factor heavily breaks scale invariance, the evolution of the perturbations is essentially non-adiabatic, and eq.~\Eq{spec} is in significant conflict with observations. If anisotropic scaling correctly defines the UV behaviour of nature, this result indicates that either a description of inflationary physics with a classical metric may be inadequate, or that the proposed scalar sector should be revised in one or more of its components. As an example, if one changed sign of the potential term in eq.~\Eq{sca1} (while keeping $\b_2>0$ in order to recover the correct IR limit), also large-wavelength scalar perturbations would be scale invariant.


\section{Discussion}

With respect to other theories defined on the same manifold, models satisfying the principle of detailed balance generally have simpler renormalization properties, the reason being that they are partly determined by the lower-dimensional theory with action $W$ (see \cite{Zin86,ZZ,Hor08} for examples). It will be interesting to study the renormalization group flow of the theory considered in \cite{Hor2,Hor3} and herein, in order to clarify its UV behaviour and make a more precise comparison with other candidates of quantum gravity. This should also clarify the consequences of the `separate detailed balance' assumption made in section \ref{acti}, eqs.~\Eq{sb1} and \Eq{sb2}.

Ho\v{r}ava theory of gravity is a concrete theoretical framework within which to embed trans-Planckian phenomenological models of inflation. As the scalar spectrum is scale dependent and observationally unviable, one might question the assumption that inflation admits a perturbative semi-classical formulation as in standard general relativity. A study of the early universe in the full quantum theory could clarify this issue. For instance, some traits of the quantum dynamics are captured once the running of the couplings is taken into account and the equations of motion are RG improved, like in the Planckian cosmology of asymptotically safe gravity \cite{BoR1,ReS3,BoR3}. However, the crux of the problem is the $-$ sign in front of the dominant ultraviolet correction to the scalar dispersion relation, which gives rise to the exponential factor in eq.~\Eq{spec}. This is a consequence of the `separate' detailed balance principle and an intrinsic feature of the model we proposed. Therefore one should consider to relax, modify or abandon this condition, e.g. by allowing for non-minimal couplings through a `full' detailed balance or considering an alternative definition of the scalar sector.

\bigskip

\textbf{Note added.} After the completion of our paper, we became aware of ref.~\cite{TaS}, where tensor perturbations of Ho\v{r}ava's theory on a de Sitter background are considered. Where the analyses overlap, they agree.


\begin{acknowledgments}
The author is supported by NSF Grant No. PHY0854743, the George A. and Margaret M. Downsbrough Endowment and the Eberly research funds of Penn State.
\end{acknowledgments}
 

\end{document}